\documentclass[twocolumn]{aastex62}

\definecolor{mygreen}{RGB}{34,139,34}

\graphicspath{{./}{figures/}}

\submitjournal{ApJ Letters}

\shorttitle{Mass excess in M62}
\shortauthors{Abbate et al.}
\newcommand{\msun}{\,\mathrm{M}_{\sun{}}}

\begin{document}

\title[Mass excess in M62]{Evidence of non-luminous matter in the center of M62}

\correspondingauthor{Federico Abbate}
\email{f.abbate@campus.unimib.it}

\author[0000-0002-0786-7307]{Federico Abbate}
\affiliation{Dipartimento di Fisica `G. Occhialini', Universit\`a degli Studi Milano - Bicocca, Piazza della Scienza 3, Milano, Italy}
\affiliation{INAF - Osservatorio Astronomico di Cagliari, Via della Scienza, I-09047 Selargius (CA), Italy}

\author{Andrea Possenti}
\affiliation{INAF - Osservatorio Astronomico di Cagliari, Via della Scienza, I-09047 Selargius (CA), Italy}
\affiliation{Dipartimento di Fisica, Università di Cagliari, SP Monserrato-Sestu, km 0.7, 09042 Monserrato, Italy}

\author{Monica Colpi}
\affiliation{Dipartimento di Fisica `G. Occhialini', Universit\`a degli Studi Milano - Bicocca, Piazza della Scienza 3, Milano, Italy}
\affiliation{Istituto Nazionale di Fisica Nucleare, Sezione di Milano Bicocca, Piazza della Scienza 3, Milano, Italy}
\affiliation{INAF - Osservatorio Astronomico di Brera, via Brera 28, I-20121 Milan, Italy}

\author{Mario Spera}
\affiliation{Dipartimento di Fisica `G. Occhialini', Universit\`a degli Studi Milano - Bicocca, Piazza della Scienza 3, Milano, Italy}
\affiliation{Dipartimento di Fisica e Astronomia `G. Galilei', University of Padova, Vicolo dell'Osservatorio 3, I--35122, Padova, Italy}
\affiliation{INFN, Sezione di Padova, Via Marzolo 8, I--35131, Padova, Italy}
\affiliation{Department of Physics and Astronomy, Northwestern University, Evanston, IL 60208, USA}
\affiliation{Center for Interdisciplinary Exploration and Research in Astrophysics (CIERA), Evanston, IL 60208, USA}



\begin{abstract}

Theoretical models suggest that intermediate mass black holes (IMBHs) may form and reside in the centers of globular clusters. IMBHs are still elusive to observations, but the accelerations of pulsars may bring along a unique fingerprint of their presence. In this work, we focus on the pulsars in the globular cluster M62. Using the new distance of M62 obtained from Gaia observations, we find that the measured pulsars' accelerations suggest a central excess of mass in the range $\left[1200, 6000\right]\msun$, corresponding to $\left[0.2, 1\right]$ percent of the current total mass of the cluster. Our analysis can not unambiguously discriminate between an IMBH or a system of stellar mass dark remnants of comparable total mass.

\end{abstract}



\section{Introduction}

Intermediate mass black holes (IMBHs) are black holes with mass between $10^2 - 10^5$ $\msun$. They are considered the missing link between stellar black holes and super-massive black holes and they may also be the seeds upon which super-massive black holes form \citep{Volonteri2010, Johnson2016, Latif2016, Mezcua2017}. Theoretical models suggest that IMBHs can be found in dense stellar environments such as globular clusters (GCs, \citealt{Bahcall1975}). In GCs, the high number of stellar collisions and gravitational encounters could favor the formation of IMBHs through repeated mergers of massive stars or stellar black holes segregated in the center  \citep{Miller2002,PortegiesZwart2002,Gurkan2004, PortegiesZwart2004, Giersz2015}. Alternatively, if the stellar black holes do not merge, they could form a massive sub-system in the cluster center that may mimic the effects of an IMBH \citep{Breen2013b}.

Studying the dynamics of stars and looking for signatures of accretion are two possible ways to reveal the presence of IMBHs in GCs.
Both methods have been extensively applied to many Galactic GCs finding upper limits or tentative detections \citep{Mclaughlin2006, Noyola2008,vanderMarel2010, Lutzgendorf2013} that have been partially disputed \citep{Baumgardt2017, Tremou2018}. Evidence of an IMBH in an  extra-galactic stellar cluster might have been found thanks to observations of a tidal disruption event \citep{Lin2018}. 

Additional evidence towards the presence of an IMBH in a GC can come from the accelerations of the cluster's pulsars, which can be measured thanks to the Doppler effect and the very stable periodic emission \citep{Perera2017a,Perera2017b,Kiziltan2017a,Kiziltan2017b,Abbate2018}.

In this paper, we focus on the GC M62, also known as NGC 6266. Using the observed velocity dispersion and surface brightness profiles, \cite{Mcnamara2012} could not exclude the presence of an IMBH while \cite{Lutzgendorf2013} showed that this cluster may contain an IMBH with mass 2000 $\msun$.  The latter  claim has been recently contested by \citet{Baumgardt2019}, who showed that the observations are better matched by theoretical models that do not include an IMBH. Furthermore, in a search for radio signatures of accretion from a hypothetical central IMBH, \cite{Tremou2018} failed in the detection. Instead, they posed an upper limit to the IMBH mass at 1130 $\msun$. M62 contains 6 known millisecond pulsars, all of which are in binary systems \citep{D'amico2001,Possenti2003,Lynch2012}. Three pulsars are located very close to the center of the cluster ($\sim 0.1$ pc). As such, they may bring along crucial information about the possible presence of a central IMBH.
In this work, we use the accelerations of the pulsars in M62 to look for deviations from the published density profile of the cluster and we investigate whether this deviation can be attributed to an IMBH or a central non-luminous system.

\section{Methods} \label{sec:methods}

Millisecond pulsars are excellent tools to probe the dynamics of M62 thanks to their very stable rotation. The high stability allows us a high-accuracy measure of the rotational period derivatives which, in turn, are linked to the pulsars accelerations (and derivatives) by the Doppler effect. Furthermore, due to mass segregation \citep{Spitzer1987}, pulsars are usually located close to the center of the cluster giving us insights into the mass distribution in the cluster's innermost regions.

The line-of-sight acceleration  of a pulsar is related to its rotational period derivative $\dot P$ by the equation
\begin{equation}
    \left( \frac{\dot P}{P}\right)_{\rm meas} = \left( \frac{\dot P}{P}\right)_{\rm int} + \frac{a_c}{c} +\frac{a_g}{c} -\frac{\mu^2 D}{c},
\end{equation}
where
$({\dot P}/P)_{\rm\, int}$ is the spin down caused by magnetic dipole braking, ${a_c}/{c}$ is the acceleration along the line of sight caused by the gravitational potential of the GC, ${a_g}/{c}$ is the relative acceleration of the cluster with respect to the Solar System in the gravitational potential of the Galaxy, ${\mu^2 D}/{c}$ is the centrifugal acceleration caused by the proper motion of the pulsars called Skhlovskii effect \citep{Shklovskii1970}, $\mu$ is the proper motion of the pulsar, $D$ is the distance of the cluster, and $c$ is the speed of light.

The contribution of the Shklovskii effect for the pulsars in M62 can be measured assuming the proper motion of the GC ($\mu = 5.79$ mas yr$^{-1}$, \citealt{Gaia2018}) multiplied by $c$. This means we get a typical acceleration of $\sim 10^{-10}$ m s$^{-2}$. The contribution of the gravitational potential of the Galaxy at the  GC's Galactic coordinates ($l_{\rm g}=353.574^{\circ}$, $b_{\rm g}=7.318^{\circ}$), derived from the Galactic potential model of \citet{Kuijken1989} is \citep{Nice1995}:

\begin{equation}
a_g= -\cos(b_{\rm g}) \left( \frac{\Theta_0^2}{R_0}\right) \left[ \cos(l_{\rm g}) + \frac{\beta}{\sin^2(l_{\rm g})+\beta^2}\right] 
\end{equation}
where $R_0= 8178 \pm 12_{\rm stat} \pm 22_{\rm sys}$ kpc is the distance of the Galactic center \citep{Gravity2019}, $\Theta_0 = 240 \pm 8$ km s$^{-1}$ \citep{Sharma2014} and $\beta= (D/R_0)\cos(b_{\rm g})-\cos(l_{\rm g})$. We obtain $a_g \simeq 5\times 10^{-10}$ m s$^{-2}$.  

The intrinsic spin down caused by magnetic dipole braking can vary from pulsar to pulsar. The average value can be estimated from the statistical distribution of intrinsic spin down from the population of MSPs from the Galactic disk. For them the observed $({\dot P}/{P})_{\rm meas}$ is dominated by the intrinsic spin down.
Taking into account the population of Galactic MSPs \citep{Manchester2005}, \cite{Abbate2018} find that the contribution on the acceleration due to the intrinsic spin down is of the order of $10^{-9}$ m s$^{-2}$.

The measured acceleration, $c( {\dot P}/{P})_{\rm meas} $, of the pulsars in M62 in absolute value varies from $3 \times 10^{-9}$ m s$^{-2}$ for pulsar C to $3 \times 10^{-8}$ m s$^{-2}$ for pulsar B. This means that the dominant contribution is due to the acceleration caused by the gravitational potential of the cluster, $a_c$. This acceleration can be estimated directly using formulas derived from the density distribution of the cluster, usually considered to be a King profile \citep{King1962}. These equations, however, depend on the distance of the pulsar from the center of the cluster along the line of sight. This quantity is not directly measurable from the observations and is therefore an unknown. 

To calculate analyitically $a_c$ and search for the gravitational effects of an IMBH we have to fit for the structural parameters of the GC together with the line-of-sight distances of the pulsars. We use a Markov Chain Monte Carlo (MCMC) code first introduced in \cite{Prager2017} and further developed in \cite{Abbate2018}. This code uses the information derived from the position of pulsars in the plane of the sky and their acceleration along the line of sight to find the best match for (i) the  structural parameters of the cluster, (ii)the position along the line of sight of each pulsar, and (iii) the mass of a dark massive object, M$_{\bullet}$, which corresponds to an IMBH if M$_{\bullet} >10^2 \msun$. As structural parameters, we use the core radius $r_c$ and the central mass density $\rho_0$. Alternatively to the central density, the one dimensional central velocity dispersion, $\sigma_c$, can be used. The latter has the advantage of being measurable in observations and is linked to the other two parameters by the equation \citep{Binney2008}:
\begin{equation}
    \sigma_c =\sqrt{\frac{4\pi G \rho_0 r_c^2}{9}}.
\end{equation}
An additional parameter that is used by the MCMC code is the mass segregation parameter $\alpha$. This is linked to the column number density profile of the pulsars by the following formula \cite{Lugger1995}:

\begin{equation} \label{dens_perp}
n(x_{\perp})= n_0 \left(1 + x_{\perp}^2 \right)^{\alpha/2},
\end{equation} 
where $n_0$ is the central number density, and $x_{\perp}$ is the distance from the center in the plane of the sky in units of core radii, defined as $x_{\perp}=R_{\perp}/r_c$. The mass segregation parameter $\alpha$ is related to the mass of the pulsars by the formula $\alpha= 1- 3q$, where $q$ is the ratio between the mass of the pulsar $M_{\rm p}$ and the mass of the stellar population that dominates the dynamics in the cluster ($q= \rm{M_{\rm p}/M_{*}}$). If pulsars have the same mass as the dominant mass class we obtain $\alpha=-2$, which is the value for the single-mass analytical King model \citep{King1962}. In typical GCs $\alpha \simeq -3$ \citep{Prager2017, Abbate2018}. We take the prior on this parameter to be a Gaussian centered on $-3$ with a dispersion of $0.5$.

We use the $\alpha$ parameter to estimate the probability that a pulsar will have a given line-of-sight distance from the center of the cluster. Furthermore, we assume that the GC is spherical and that the distribution of pulsars along the line of sight  is the same as the number density profile along the directions on the plane of the sky.

\section{observational parameters} \label{section:optical}

The GC M62 has been observed at different wavelengths over the years in order to measure the parameters of the cluster. The properties important for the present work are shown in Table \ref{tab:obs_properties}.

    \begin{table}
        \centering
        \begin{tabular}{ccc}
            Parameter & Value & Reference \\
            \hline
             Stellar mass & & \\
             & 6.74 $\pm$ 0.05 $\times 10^5$ $\msun$ & 1 \\
             \hline
             Distance & & \\
             & 6400 $\pm$ 180 pc& 2 \\
             \hline
             Core radius & &\\
              & 0.36 pc& 1\\
             \hline
             Velocity dispersion & & \\
             Radial & 15.0 $\pm$ 1.1 km s$^{-1}$&  3\\
             Proper motion & 15.3 $\pm$ 0.4 km s$^{-1}$ & 4\\
             \hline
        \end{tabular}
        \caption{ Properties of the GC M62 using optical data. 1, \protect\cite{Baumgardt2018}; 2, \protect\cite{Baumgardt2019}; 
        3, \protect\cite{Kamann2018}; 4, \protect\cite{Watkins2015}.}
        \label{tab:obs_properties}
    \end{table}

The most recent distance estimate obtained through Gaia observations \citep{Baumgardt2019} is $6400 \pm 180$ pc. This value is smaller than the value found in \cite{Harris1996} (2010 edition) which was 6800 pc. Throughout the paper we will use the new distance estimated by \citet{Baumgardt2019}.

The core radius can be derived from the luminosity density profile with a single mass King model \citep{King1962}. Using Hubble Space Telescope observations, \cite{Miocchi2013} found $15.4 \pm 0.6$ arcsec, which corresponds to $0.48 \pm 0.02$ pc. In contrast, recent N-body simulations that match the surface density, the velocity dispersion and the mass function of M62, suggest a smaller core radius of about 0.36 pc \citep{Baumgardt2018}\footnote{For a complete list of structural parameters derived from N-body simulations for all known GCs see: https://people.smp.uq.edu.au/HolgerBaumgardt/globular/\\parameter.html}.

Using radial velocity measurements, \citet{Kamann2018} obtained a value of the central velocity dispersion of $15.0 \pm 1.1$ km s$^{-1}$. Using proper motion data, \citet{Watkins2015} estimated a value of  $0.504 \pm 0.004$ mas yr$^{-1}$, corresponding to which converted to km s$^{-1}$ using the newly determined distance becomes $15.3 \pm 0.4$ km s$^{-1}$.

Further information about the dynamical structure of the cluster can be gathered from radio observations of the pulsars. The ephemerides of the pulsars in M62 are taken from \cite{Lynch2012} and include information on the position of the pulsars in the plane of the sky, their rotational periods (and higher time derivatives), and the binary parameters. Using the position of the center of gravity as given in \cite{Miocchi2013}, pulsars B, E and F are  at about 4 arcseconds from the cluster's center. At a distance of 6400 pc \citep{Baumgardt2017} this corresponds to about 0.1 pc. This distance is comparable with that of the innermost bin in the velocity dispersion profile presented by \cite{Watkins2015}. The accelerations of these three pulsars are crucial to get insights into the central cluster's mass distribution.

\section{Results}

\begin{figure} 
\centering
\includegraphics[width=\columnwidth]{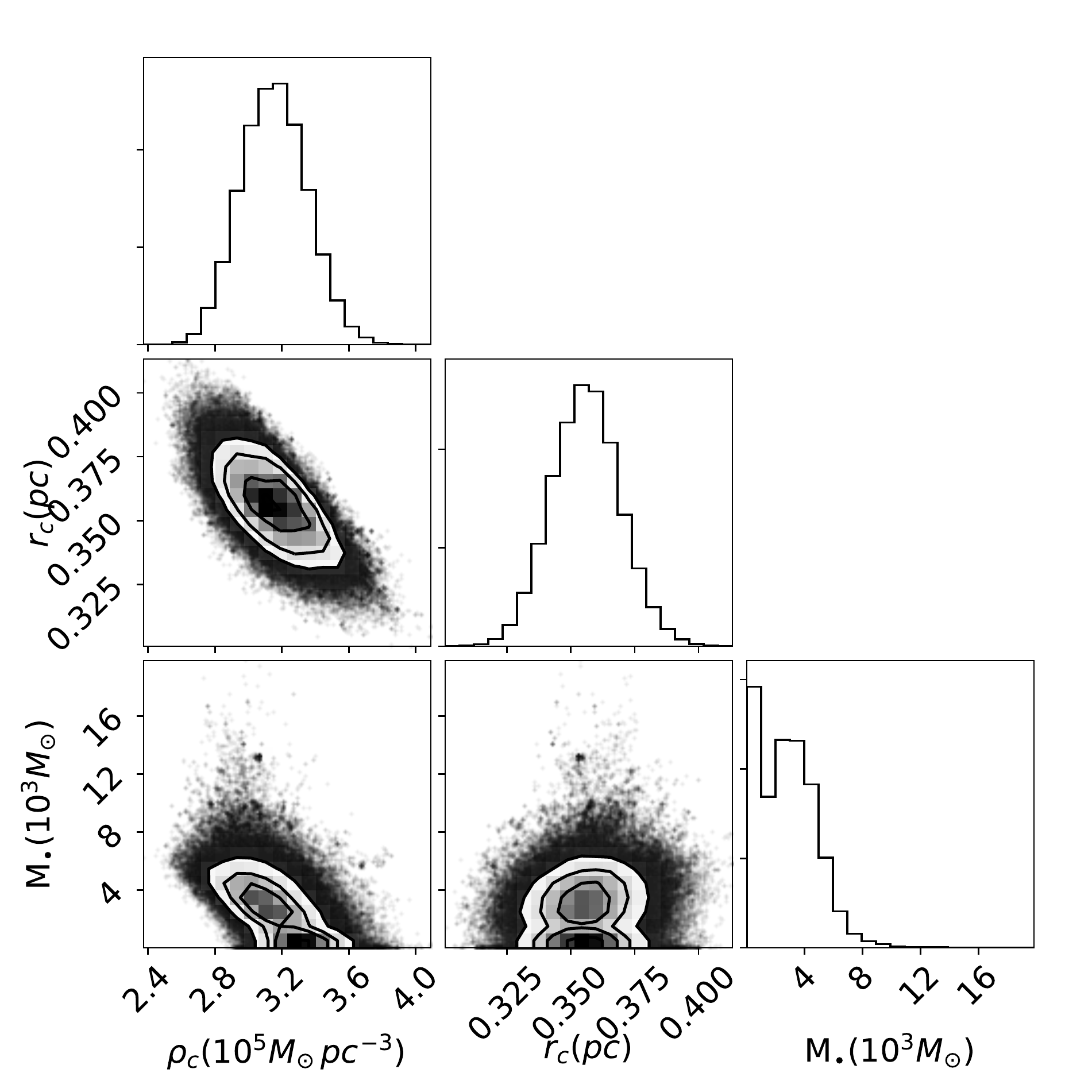}
\caption{Corner plot showing the posterior distribution of the central density, the core radius and the mass of the central dark object.}
\label{triangle_baumgardt}
\end{figure}

\begin{figure} 
\centering
\includegraphics[width=\columnwidth]{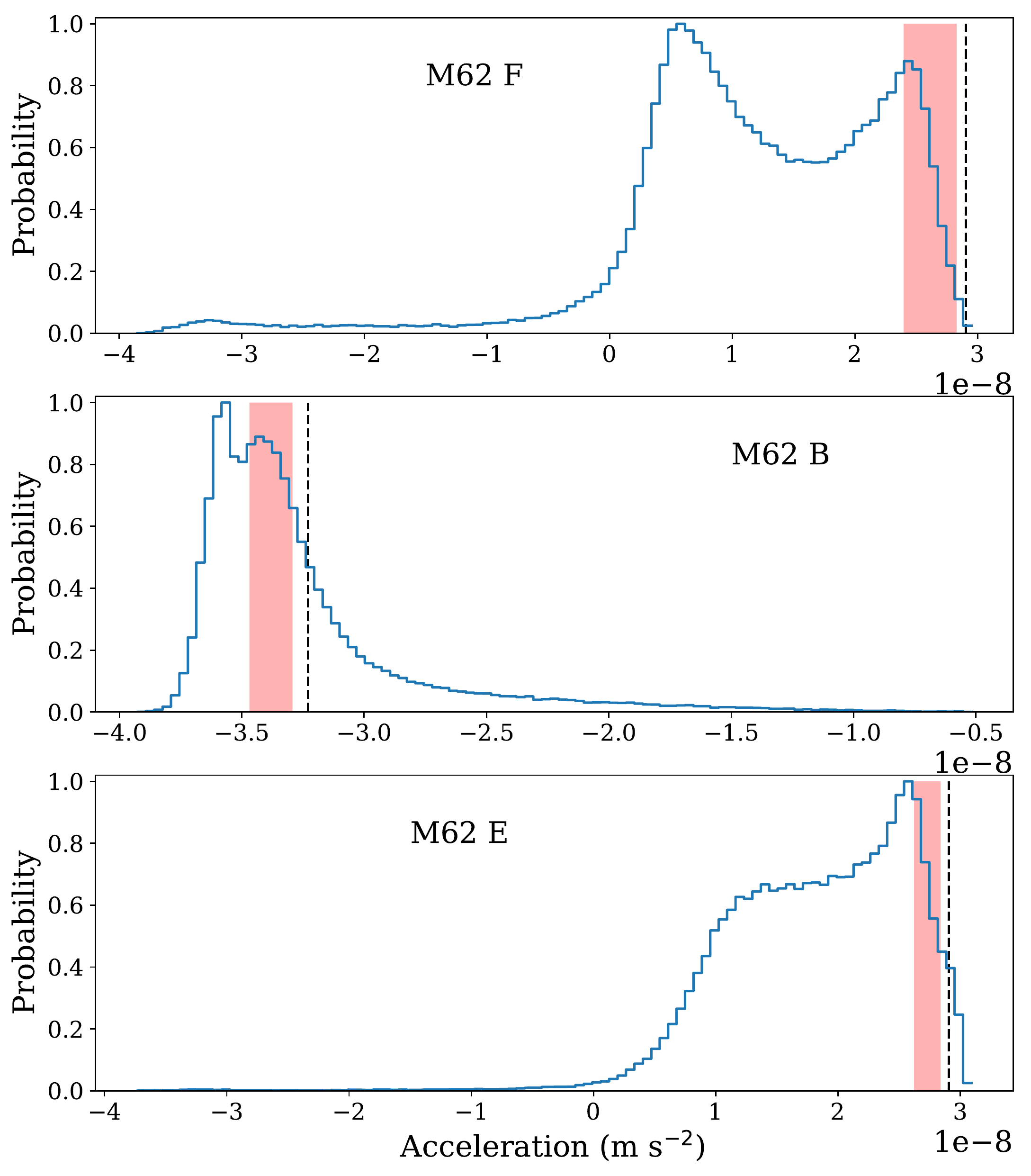}
\caption{Posterior distribution for the accelerations of the pulsars F, B and E, the three pulsars closer to the center. The dashed line shows the measured value of $\dot P/P$ and the red shaded area is the 1$\sigma$ interval of the cluster acceleration after removing the Galactic contribution, the Shkolvskii effect and the intrinsic spin-down.}
\label{pulsars_acc}
\end{figure}

\begin{figure} 
\centering
\includegraphics[width=\columnwidth]{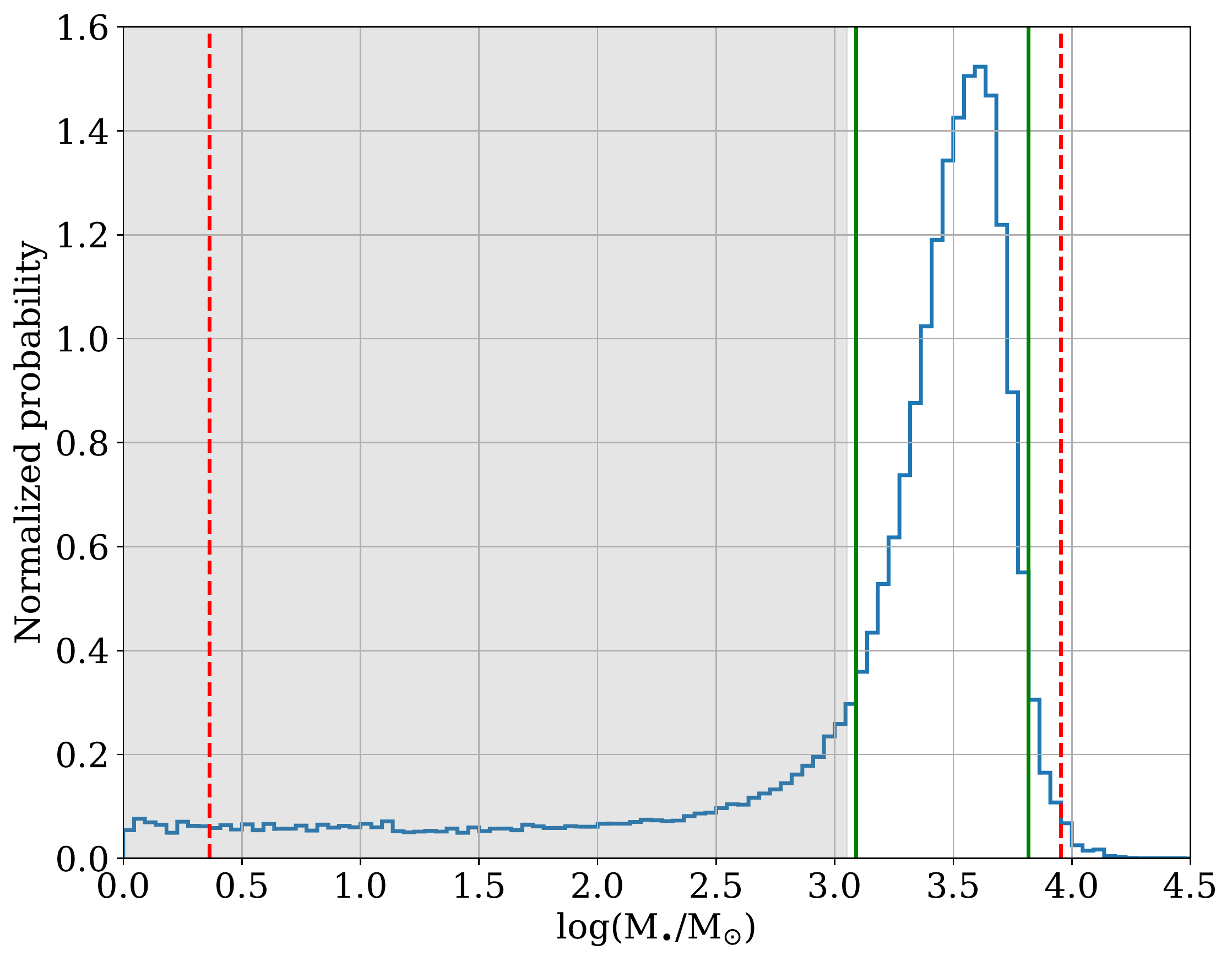}
\caption{Posterior distribution function on the mass of the central dark object in logarithmic units. The green lines show the 68 percent interval and the dashed red lines show the 95 percent interval. The shaded interval is the allowed IMBH mass range from \protect{\cite{Tremou2018}}. We assume a core radius of 0.36 pc as in \protect{\cite{Baumgardt2018}}. }
\label{BH_baumgardt_fixed}
\end{figure}

To perform the fit for the IMBH we need to assume some priors on the parameters. We use the most precise values determined from optical observations (see Section \ref{section:optical}). We take the velocity dispersion measured with proper motion data and the core radius as estimated in \cite{Baumgardt2018} and the distance measured with Gaia data. The priors for the parameters are gaussians centered around those values and with the quoted uncertainty as standard deviation. For the core radius the standard deviation was chosen to be 0.01 pc (the same 
fractional uncertainty as the distance).
The corner plot of the core radius, central density and central dark object mass is shown in Figure \ref{triangle_baumgardt}. 
The comparison between the measured accelerations of the pulsars and the ones predicted by the best-fitting model are shown in Figure \ref{pulsars_acc}. Only the innermost three pulsars are shown as they are the most affected by the presence of the central object.
The posterior distribution function for the mass of the central dark object in logarithmic units is shown in Figure \ref{BH_baumgardt_fixed}. We find that an excess mass of 3900 $\msun$ is needed in the center to explain the measured pulsar accelerations. The 68 percent interval is $(1200,6600)$ $\msun$. The posterior distribution becomes compatible with the case of no IMBH (M$_{\bullet}<10^2 \msun$) if we consider the 95 percent interval. This excess is on top of the mass distribution estimated for a single-mass King model derived from the observed optical parameters and is located within 0.2 pc (the distance from the center of the innermost pulsar, F).
The central mass density converges to $(3.1 \pm 0.2) \times 10^5$ $\msun$ pc $^{-3}$. This value is almost twice than what estimated through N-body simulations in \cite{Baumgardt2018}, which is $1.6\times 10^5$ $\msun$ pc $^{-3}$. This means that an excess of mass in the center is needed to explain the accelerations of the pulsars.
This is apparent also in the lower left panel of the corner plot in Figure \ref{triangle_baumgardt} where we see that if there is no IMBH, the central density must increase in order to compensate for the missing mass. 

To test the nature of the excess of mass, we measure the central mass-to-light ratio. First, we transform the central density in a surface mass density by multiplying it by $2 r_c$ \citep{Freire2005}. This is compared with the surface brightness density measured in \cite{Noyola2006} and converted into Solar luminosities per square parsec. The resulting central mass-to-light ratio and the 68 percent interval is $6.4_{-0.7}^{+2.1}$ $\msun \,\mathrm{L}_{\sun{}}^{-1}$.
This value is much higher than what is seen at the cluster centers in N-body simulations of different GCs (see Fig. 3 in \citealt{Baumgardt2017}).
Alternatively to an IMBH, the central excess of mass could be explained with a system of stellar dark remnants like massive white dwarfs, neutron stars or black holes. A system of this kind could form in the first evolutionary phases of a GCs and survive up until now if the host cluster has a sufficiently long half-mass relaxation time like M62 ($\sim 1$ Gyr) \citep{Breen2013,ArcaSedda2016,ArcaSedda2018}.

\subsection{Uncertainties in the optical parameters} 
\label{section:uncertainties}

\begin{figure} 
\centering
\includegraphics[width=\columnwidth]{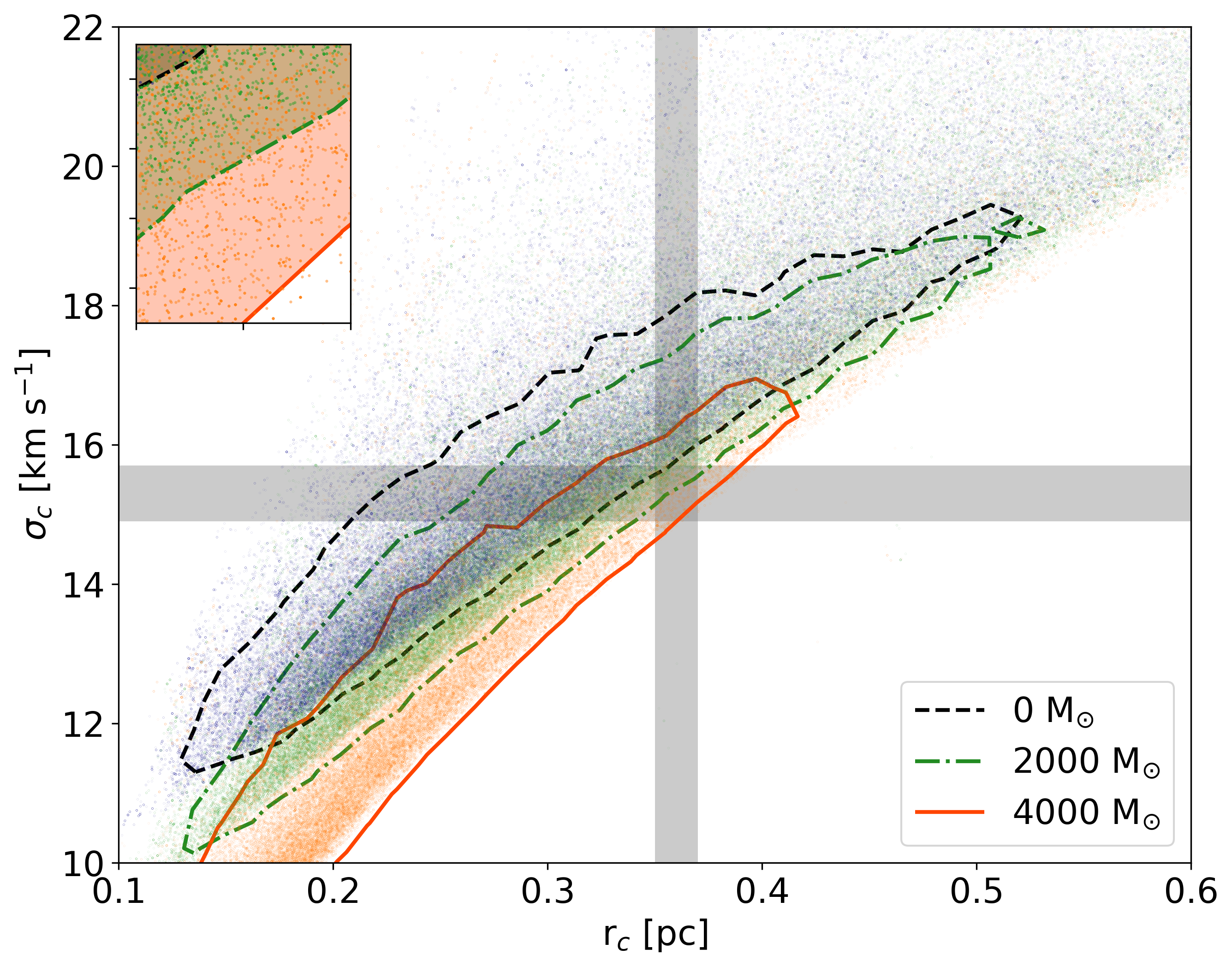}
\caption{Contour plot of the posterior distribution of core radius and velocity dispersion as resulting from the MCMC fit. The shaded grey area shows the 1$\sigma$ interval of the best determinations of core radius and velocity dispersion. The different sets represent fits with three values of IMBH mass: 0 $\msun$ (grey), 2000 $\msun$ (green), 4000 $\msun$ (orange). The coloured lines show the 68 percent intervals of these distributions. The inset shows a zoom on the central shaded area.}
\label{Contour_plot}
\end{figure}

We have shown that the probability to find a central excess of mass in M62 is quite high. To check that our finding is not affected by possible systematic errors on the values of the core radius and on the central velocity dispersion, we repeat the MCMC fit by setting the IMBH mass to fixed values, without assuming any priors on the core radius and on the velocity dispersion. Figure \ref{Contour_plot} shows the posterior distribution of the fits for three masses of IMBH, 0 $\msun$, 2000 $\msun$ and 4000 $\msun$. In the same plot we show, with the shaded area, the observed 1$\sigma$ range for the core radius and velocity dispersion. In the case with no IMBH the shaded interval intersects the 68 percent interval of the posterior distribution only in a very small region (shown in the inset) but, as the mass of the IMBH increases, the region of intersection grows. However, if the real core radius is smaller than the measured one it could be possible to explain the observed pulsar accelerations without the need of an IMBH. The same is true if the true velocity dispersion is higher than what is measured. It is important to note that also in these cases where the IMBH is not needed, the mass to light ratio must remain at the same values.

To test the relation between the mass-excess and the
core radius and velocity dispersion of the cluster,
we run the MCMC code, using two mock GCs; the first (second)  with a core radius of 0.16 pc (0.58 pc) and a total mass of $9 \times 10^4$ $\msun$ ($8.4 \times 10^5$ $\msun$) simulated in \citep{Abbate2019}.  The simulations host IMBHs of different masses in the center. Here, we show the result of the smallest cluster in the case there is no IMBH in the core. We first run the code with a prior on $r_c$ corresponding to the true value of 0.16 pc inferred from the simulated GC, and later with inject an erroneous value of $r_c$ overestimated by 60 percent to 0.26 pc. This overestimation corresponds to the difference between the largest measured $r_c$ for M62, 0.48 pc \citep{Miocchi2013}, and the value it should have to be compatible with no IMBH keeping the same velocity dispersion taken from Figure \ref{Contour_plot}, that is $\sim 0.3$ pc. 
We extract six pulsars (using $\alpha= 3$), and measure their acceleration along the line of sight for the two cases.

\begin{figure} 
\centering
\includegraphics[width=\columnwidth]{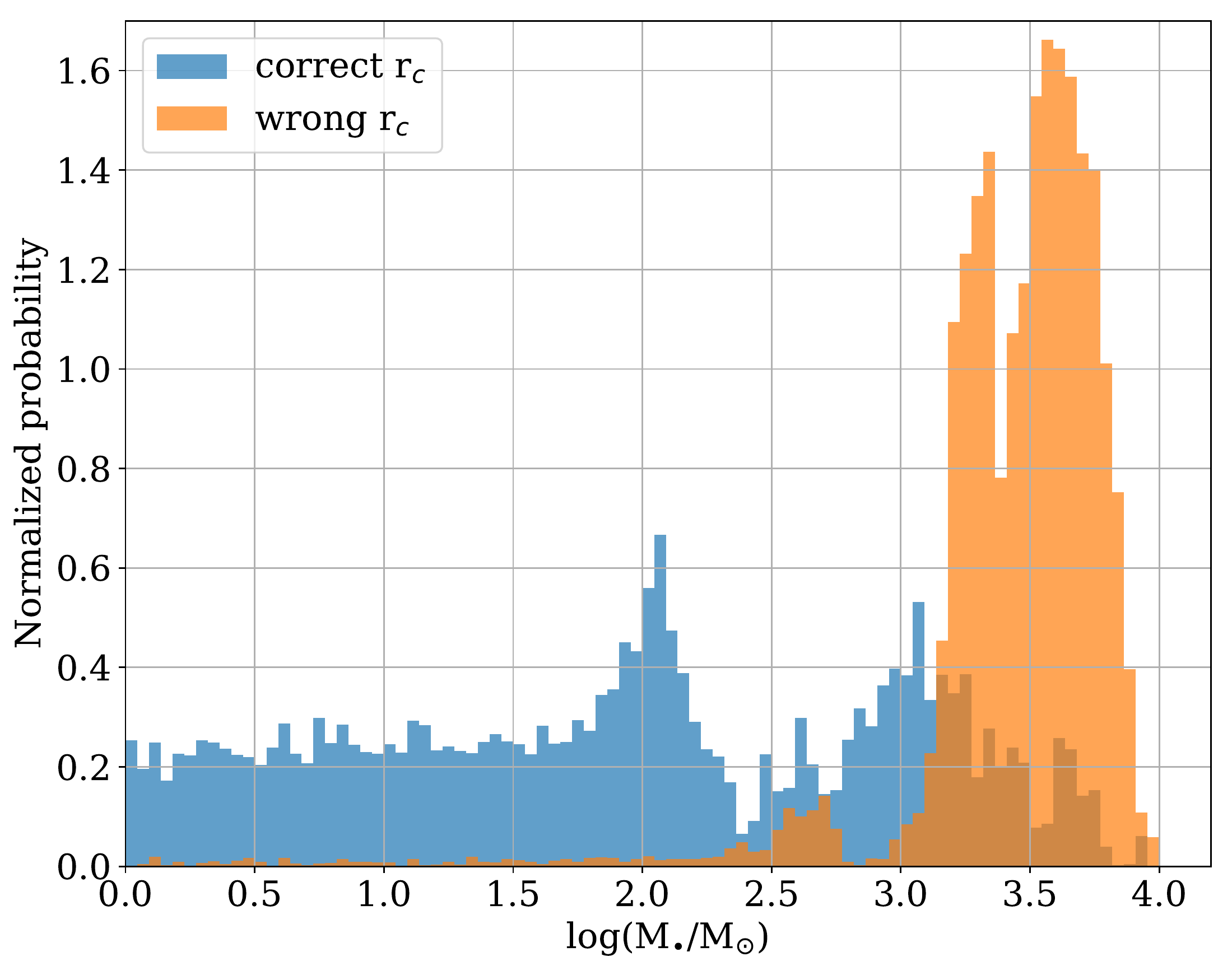}
\caption{
Posterior distributions of the mass of the central dark object resulting from a simulated GC with no such mass excess. Assuming the correct value of core radius (blue histogram), the distribution is consistent with a null mass excess. Instead, assuming a core radius overestimated by 60 percent (orange histogram), the distribution indicates a mass excess of $\sim3500$ $\msun$ which is not present in the simulation.}
\label{simulations_0bh}
\end{figure}

Figure \ref{simulations_0bh} shows the histograms of the posterior distributions of the central dark object mass from the mock GC. The distribution in orange, corresponding to the incorrect value of $r_c,$ shows a peak at $\sim 3500$ $\msun$ which is incompatible with 0 $\msun$ at the 3$\sigma$ level, whereas
the distribution in blue is consistent with the correct assumption of no-IMBH, at 1$\sigma$ level.

We repeat the test using the simulation of the same GC with an IMBH of 1000 $\msun$ at the center and for the more massive GC with a core radius of 0.58 pc. The simulations of both clusters strongly support that an error in the determination of the cluster parameters can lead to an incorrect value of the mass of the IMBH.

\section{Discussion}

We have used the measured accelerations of the pulsars in M62 and the new distance estimate by Gaia to predict the existence of an excess of mass located within the central 4 arcseconds of M62. The existence of a concentration of non-luminous mass is confirmed by looking at the very high central mass-to-light ratio ($\sim 6$). The source of this dark mass excess could be a single IMBH of $\sim 3900$ $\msun$ or a system of massive dark remnants segregated in the center of similar total mass. This result is obtained strictly assuming the nominal values of the core radius and velocity dispersion from the latest optical observations \citep{Watkins2015, Baumgardt2018}. 
We tested the dependence of the result from these assumptions and found important correlations between the aforementioned optical parameters and the mass of the IMBH resulting from the fit. Assuming a smaller core radius or a faster velocity dispersion would result in an IMBH of smaller mass. 

We tested the correlation using simulated GCs of known parameters and we confirmed that an alternate assumption on the core radius can mimic the necessity of an IMBH nor exclude its presence. Thus, more precise estimates of these parameters are needed to uniquely confirm the presence of an IMBH in the cluster.

The optical observations of M62 are compatible with an IMBH of a few thousands $\msun$ \citep{Mcnamara2012} with a tentative claim of $\sim 2000$ $\msun$ \citep{Lutzgendorf2013} which has been disputed in \cite{Baumgardt2017}. According to searches for accretion signatures \citep{Tremou2018}, there is an upper limit on the mass of the IMBH of 1130 $\msun$. From the results of our analysis, we cannot either confirm these claims or reduce the upper limits but, using the published optical parameters of M62, we have found that $\sim 4000$ $\msun$ of low luminosity matter must be contained within the central 0.2 pc of the cluster. This `dark' matter can be composed of an unconstrained ratio between one or more IMBHs or a large number of non luminous remnants of stellar origin.

Present-day clusters, with half mass relaxation times $\gtrsim 1$ Gyr, may have retained many stellar mass dark remnants in their centers \citep{Breen2013,ArcaSedda2018}. Such sub-systems can reach masses of 1000 - 10000 $\msun$ for GCs born very massive \citep{ArcaSedda2016}. 
\cite{ArcaSedda2019} showed that M62-like GCs build up a sub-system of dark remnants in their center in about half of the authors' N-body realizations. Though highly speculative, some of these stellar remnants might be in the form of binary black holes, that are now observed as powerful gravitational wave sources \citep{Ligo2018}.
Furthermore, \cite{Ye2019} showed that the presence of a large number of black holes in the center of the clusters should drive out all of the MSPs found inside and would be incompatible with the observed set of pulsars. The stability of this system and the coexistence of stellar black holes and MSPs should be studied with dynamical simulations focused on this cluster.

Further observations of the cluster are necessary to find the source of the excess of mass. Optical observations will help to determine the structural parameters with higher precision and radio observations will better constrain the accelerations of the pulsars and to look for new pulsars possibly even closer to the center.


\section*{Acknowledgements}
This work has been funded using resources from the research grant ``iPeska'' (P.I. Andrea Possenti) funded under the INAF national call Prin-SKA/CTA approved with the Presidential Decree 70/2016. We acknowledge the ``Accordo Quadro INAF-CINECA (2017)'' and the CINECA-INFN agreement for the availability of high performance computing resources and support. MC acknloedges finantial support 
by INFN under the grant Teongrav. MS acknowledges funding from the European Union's Horizon 2020 research and innovation programme under the Marie-Sklodowska-Curie grant agreement No. 794393.

\bibliography{IMBH_biblio.bib} 



\end{document}